# REVIEW

# Feedback in low-mass galaxies in the early Universe

Dawn K. Erb[1]

The formation, evolution and death of massive stars release large quantities of energy and momentum into the gas surrounding the sites of star formation. This process, generically termed 'feedback', inhibits further star formation either by removing gas from the galaxy, or by heating it to temperatures that are too high to form new stars. Observations reveal feedback in the form of galactic-scale outflows of gas in galaxies with high rates of star formation, especially in the early Universe. Feedback in faint, low-mass galaxies probably facilitated the escape of ionizing radiation from galaxies when the Universe was about 500 million years old, so that the hydrogen between galaxies changed from neutral to ionized—the last major phase transition in the Universe.

Although feedback from star formation is a complex phenomenon involving many processes over a wide range of physical scales, its most basic consequences are not difficult to understand: stars form out of gas, and therefore any process which either removes gas from a galaxy or prevents gas from condensing into new stars will have an important effect on the subsequent evolution of that galaxy. The topic of galactic-scale outflows in galaxies has been of interest to astronomers for half a century at least[1], but over the past two decades, the explosion of new facilities enabling large surveys of galaxies in the early Universe and the increasing sophistication of simulations of galaxy evolution have led to widespread recognition that outflows in galaxies at early times are both ubiquitous and essential: evidence for outflows is seen in nearly all star-forming galaxies at high redshifts[2–7], and without feedback, gas in galaxies would cool and form too many stars, resulting in much higher stellar masses than we observe today.

As the sizes of galaxy surveys increase and as new, more sensitive instruments are developed, we are beginning to constrain the properties and relative abundance of faint, low-mass galaxies in the early Universe. It is an opportune time to emphasize the importance of this population of galaxies: a new generation of sensitive, multi-object near-infrared spectrographs is offering access to their rest-frame optical emission lines (key to constraining the physical conditions in their star-forming regions) for the first time, while a variety of studies are pointing out the substantial contribution of such faint, low-mass galaxies to both the overall star-formation density of the Universe and the reionization of the Universe, when ionizing radiation from the first generation of stars and galaxies reionized the hydrogen gas in the intergalactic medium, which had been neutral since protons and electrons first combined 375,000 years after the Big Bang.

Complex, multi-phase galactic outflows are likely to determine the baryon and heavy-element content of both galaxies and the intergalactic medium. Outflows are seen in galaxies with unusually high star-formation rates in the local Universe, but the rest-frame ultraviolet spectra of galaxies at high redshifts show that outflows are a generic feature of star-forming galaxies in the early Universe. However, we have few constraints on the properties of feedback in the faint, low-mass (here defined as galaxies with stellar masses $\lesssim 10^9 M_\odot$) end of the distant galaxy population. Such constraints are badly needed: since outflow properties in the local Universe are observed to scale with mass and star-formation rate, an increase in the dynamic range over which these properties are measured at high redshifts will improve our understanding of the physical mechanisms behind feedback. Faint, low-mass galaxies are now also recognized as the likely source of many of the ionizing photons responsible for reionization, the last major phase transition in the Universe, and it is likely that feedback processes aided the escape of these photons. Current constraints on feedback in low-mass galaxies come mostly from Lyman α (Lyα) emission, which can indicate the presence of outflows but is unlikely to map directly to outflow velocity. However, future facilities such as the James Webb Space Telescope and the 30-m-class ground-based telescopes will provide a much more detailed view of this key process in low-mass galaxies.

## An overview of galactic outflows

Both locally and in the distant Universe, galaxies with intense star formation concentrated in a small volume exhibit dramatic outflows of gas. The prototypical example of such a galaxy is the local starburst M82, shown in Fig. 1. M82 is a nearby disk galaxy with a powerful central starburst, thought to be triggered by interaction with its companion galaxy M81[9,10]. The nearly edge-on orientation of M82's disk reveals an extended, bi-conical outflow emerging perpendicular to the disk of the galaxy and emanating from its central starburst. Detailed, multi-wavelength observations of this outflow reveal its complex, multi-phase nature: hard X-ray emission (X-rays with energies $\gtrsim 10$ keV) traces hot gas with a temperature of 30–80 million K and terminal velocity of about 2,000 km s$^{-1}$, well above M82's escape velocity of about 460 km s$^{-1}$ (ref. 11); softer X-rays and Hα emission from ionized gas indicate velocities of 600–800 km s$^{-1}$ and suggest the presence of shocks[11,12]; and observations of CO emission indicate that the outflow also contains at least $3 \times 10^8 M_\odot$ of molecular gas, reaching a maximum velocity of about 230 km s$^{-1}$ (ref. 13). As in other such galaxies, the mass of gas in the outflow is estimated to be comparable to the mass of gas being formed into stars[14]. M82 is thus the archetypal example of a galaxy in which the energy and momentum of star formation is heating gas to high temperatures and propelling at least some of it to velocities sufficient for the gas to escape the galaxy altogether. Since stars form out of gas, such outflows must have important consequences for the regulation of star formation in galaxies.

Although starburst galaxies such as M82 are relatively rare in the local Universe, the existence of galactic outflows in most star-forming galaxies at redshifts $z \approx 2$ and higher is now well-established[2,3,5,6]. It is also now clear that the star-formation-rate density of the Universe peaks at $z \approx 2$–3 (ref. 15); star-formation feedback is therefore an essential part of the evolution of galaxies during the peak epoch of star formation, with a variety of important consequences for the properties of galaxies and the intergalactic medium. These consequences include the shape of the relationship between the heavy-element content of galaxies and their mass; in both local and high-redshift galaxies, more massive galaxies are more chemically enriched[16–18]. Elements heavier than helium (termed 'metals' by astronomers) are products of star formation and are released into the interstellar medium as stars evolve and die, enriching the gas and the next generation of stars. However, galactic outflows may drive some of these metals out of the galaxy instead, decreasing the overall metallicity and substantially modulating the relationship between metallicity and stellar mass[19–22].

[1]Center for Gravitation, Cosmology and Astrophysics, Department of Physics, University of Wisconsin Milwaukee, 3135 North Maryland Avenue, Milwaukee, Wisconsin 53211, USA.

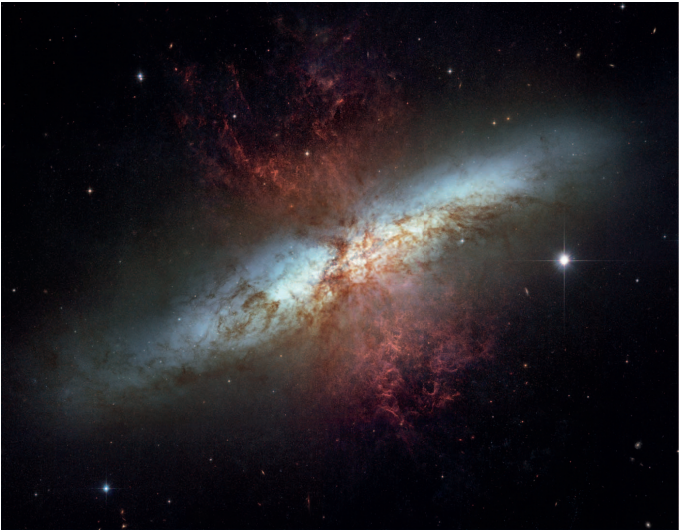

**Figure 1 | The local starburst galaxy M82 observed with the Hubble Space Telescope.** Outflowing gas, seen in emission from ionized gas shown in red, forms a bi-conical structure centred on the intense starburst in the centre of the galactic disk. Image from NASA, ESA and The Hubble Heritage Team (STScI/AURA); http://hubblesite.org/newscenter/archive/releases/2006/14/image/a/.

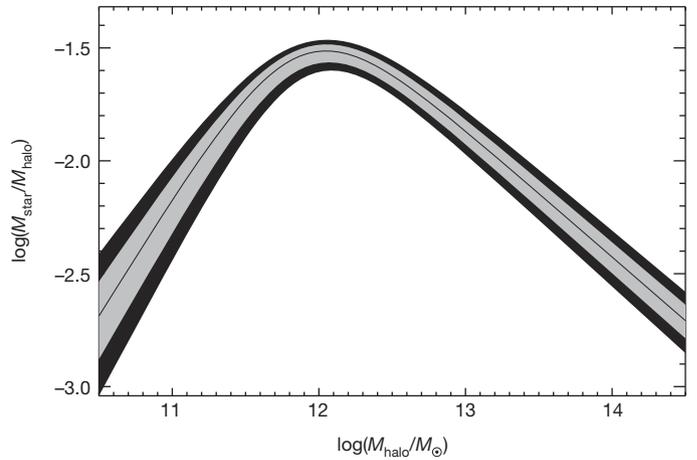

**Figure 2 | The ratio of stellar to halo masses as a function of halo mass in the local Universe.** Light and dark shaded areas show $1\sigma$ and $2\sigma$ bounds respectively. The stellar to halo mass ratio peaks at halo mass $M_{halo} \approx 10^{12} M_\odot$, and declines at both lower and higher masses. The peak and slope depend somewhat on the redshift at which the relationship is measured, but the general form holds at all redshifts studied, and indicates that the efficiency of star formation decreases at both low and high masses. Image adapted from figure 4 of ref. 34 (IOP Publishing, American Astronomical Society).

Metals thus driven out of galaxies are observed in the gas between galaxies. Most of the baryons in the Universe lie not in galaxies but in the intergalactic medium[23], and observations of the gas between and around galaxies via absorption line systems in the spectra of quasars have long shown that some of the gas is metal-enhanced[24–26]. Individual metal absorption line systems can be associated with galaxies near the line of sight to the quasar[27], while large, joint surveys of quasars and foreground galaxies have enabled the statistical study of metal absorption lines arising from gas near galaxies[24,25,28,29]. Such studies have found that metals are enhanced near galaxies to distances from each galaxy of about 180 proper kiloparsecs (kpc) at $z \approx 2.4$ (ref. 30), while in the local Universe, observations probing the gas around galaxies to distances of 160 kpc have found that cool, metal-enriched gas accounts for as much as 25%–45% of the baryons in the halo of a typical $L^\star$ galaxy[31]. While the timescales over which metals are deposited into the intergalactic medium and the mixing of metals in the gas are still uncertain[32,33], it is becoming increasingly clear that galactic outflows are key to the regulation of the baryonic content of both galaxies and the gas around them.

A further issue that must be explained by any successful model of the formation and evolution of galaxies is the fact that the stellar fraction of galaxies is not constant[34,35]: as shown in Fig. 2, the ratio of galaxy stellar mass to the mass of the host dark-matter halo reaches a maximum at halo mass $M_{halo} \approx 10^{12} M_\odot$ (with some dependence on redshift), and declines at both higher and lower masses. In other words, both high- and low-mass galaxies are less efficient at forming stars. Feedback from active galactic nuclei has typically been invoked to explain the decrease in efficiency at high masses[36,37], although recent observations indicate that fast outflows in some massive galaxies may be powered by starbursts rather than by active galactic nuclei[38,39]. At lower masses, the decreasing efficiency of star formation is likely to be due to processes associated with the formation and evolution of massive stars.

Galactic winds are driven by the energy and momentum imparted to gas by massive stars, but the relative importance of various feedback processes is still a subject of considerable study. Early work focused on the thermal energy provided by supernovae, which may drive a fast wind out of the starburst nucleus and into the galactic halo or beyond[40,41]. Many more recent models have focused on the role of momentum in driving galactic outflows; unlike the thermal energy injected by supernovae, momentum cannot be radiated away, and simple prescriptions for momentum-driven winds indicate mass outflow rates comparable to the star-formation rate and an inverse scaling of the wind efficiency (the mass outflow rate relative to the star-formation rate) with galaxy circular velocity, as suggested by observations of galactic outflows in the local Universe[42–44].

Modern models of galaxy formation and evolution emphasize the importance of feedback, but its implementation in simulations is difficult, primarily because of the range of scales involved (from cosmological scales to the scale of individual stars or at least giant molecular clouds) and the large number of complex and poorly understood physical processes. As a result, models rely on 'sub-grid' prescriptions for small-scale processes that are not resolved by the simulation. These models can reproduce observations and have demonstrated the effects of feedback on the evolution of galaxies and the intergalactic medium, but the results of these simulations may be dependent on the feedback prescriptions and how they are implemented[45,46].

Numerical simulations of galaxy formation are only now beginning to model feedback by directly modelling the physical processes involved. At present such simulations incorporate radiation pressure from starlight, energy and momentum injection from supernovae, stellar winds, and photoionization, and photoelectric heating; the results show that different feedback processes may interact in complex and nonlinear ways, and that different processes may dominate in different environments[47–49]. For example, radiation pressure probably dominates in massive, dusty and gas-rich starbursts, in which thermal energy is quickly radiated away, whereas the low gas densities of dwarf starbursts result in slow cooling times and hot winds driven by supernova heating[47]. In all cases, outflowing gas occupies a broad distribution in temperature and velocity, as seen in observations[14,50].

## Observations of outflows at high redshifts

As exemplified by M82, the archetypal starburst described above, detailed observations of galactic outflows in the local Universe reveal that they are a complex, multi-phase phenomenon, with outflows observed in hot gas traced by X-rays[51,52], ionized gas seen via optical emission lines[53], neutral gas probed by low ionization absorption lines[14,50,54], and molecular gas observed at radio wavelengths[13,55]. These observations have revealed that the properties of outflows vary with galaxy mass, star-formation rate, and the surface density of star formation: more massive galaxies with higher star-formation rates and higher star-formation-rate densities tend to drive faster outflows[14,50,56]. An observational determination of the scalings of outflow properties with galaxy properties is key to identifying the underlying mechanisms driving the outflow[42,47], but this determination is complicated by the fact that correlations appear to be relatively weak and may flatten at higher masses and star-formation rates. For example, studies find the wind velocity is proportional to star-formation rate (SFR) as follows: $v \propto \mathrm{SFR}^{\alpha}$, with $\alpha$ ranging from 0.1 to 0.35. The result is that a wide dynamic range in galaxy properties, extending to dwarf galaxies, is required to detect trends robustly[4,14,50]. Such measurements are possible in the local Universe, where the masses of the galaxies studied range from about $10^7 M_\odot$ to $10^{11} M_\odot$ and their star-formation rates range from about $10^{-2} M_\odot$ per year to $10^3 M_\odot$ per year[50,56].

At redshifts $z > 1$, observations are much more limited, owing to both the faintness of galaxies at these distances and the redshifting of key diagnostics into the near-infrared; the result is that high-redshift samples used to study outflows contain few galaxies with masses $< 10^9 M_\odot$ or star-formation rates less than a few $M_\odot$ per year. While outflows in relatively massive galaxies have been detected via observations of ionized[7] and molecular gas[57], the vast majority of our observational knowledge of feedback in galaxies at high redshift (here referring to redshifts $1 \lesssim z \lesssim 4$, at which most observations of feedback in distant galaxies have been made) comes from observations of interstellar absorption and Lyα emission lines in the rest-frame ultraviolet spectra of these galaxies. A typical example of such a spectrum is shown in Fig. 3. Because this spectrum traces the rest-frame ultraviolet, the continuum light is produced by hot, massive young stars. The strong absorption features are resonance lines from interstellar gas, blended at this resolution to include both the interstellar medium of the galaxy and gas entrained in outflows. This spectrum shows Lyα in emission, but such spectra may exhibit anything from strong emission to strong absorption, or a superposition of the two.

The first challenge that arises when studying galactic outflows from a spectrum such as this is the determination of the velocity zero point from which to measure outflow velocities; all of the strong features in this spectrum come from interstellar gas, which is not necessarily at rest with respect to the stars. Thus the systemic redshift, the redshift of the stars, must be measured from stellar features or from emission lines emanating from ionized gas surrounding the stars. Stellar features are typically too broad and weak to be measurable in spectra of distant galaxies, so observers usually measure systemic redshifts from the rest-frame optical emission lines of ionized gas, which are redshifted into the near-infrared for $z \gtrsim 1.5$. Observations in the near-infrared have historically been considerably more difficult than in the optical because of the higher sky background and the relative inefficiency of detectors, but this situation is now changing with the advent of new multi-object near-infrared spectrographs on large telescopes[58–60].

Once the systemic redshift can be determined, a characteristic pattern emerges, visible in the spectrum shown in Fig. 3: the interstellar absorption lines are blueshifted with respect to the systemic velocity, while Lyα emission, if present, is redshifted. A simple schematic of the model giving rise to this pattern is shown in Fig. 4. The blueshifted interstellar absorption lines are due to the absorption of light from the stellar continuum by foreground gas moving towards the observer in a galactic outflow (though there may also be absorption from non-outflowing gas in the interstellar medium of the galaxy), and the strength of the absorption lines depends on the covering fraction of this outflowing gas. Lyα emission is produced in H II regions surrounding the massive stars and is then strongly modified by resonant scattering: Lyα photons are absorbed and re-emitted by neutral hydrogen in and around the galaxy, with the result that they are most likely to escape the galaxy in the direction of the observer when they are backscattered from gas on the far, receding side of the outflow and thereby acquire a frequency shift that allows them to pass through the gas in the bulk of the galaxy unimpeded. This frequency shift then results in the redshift of Lyα emission relative to the systemic velocity. Galaxies at $z \approx 2$–$3$ show typical absorption line blueshifts of about $200\,\mathrm{km\,s}^{-1}$, while Lyα is typically redshifted by about $500\,\mathrm{km\,s}^{-1}$; however, measurements of individual galaxies reveal substantial variation in both quantities[2,3,5].

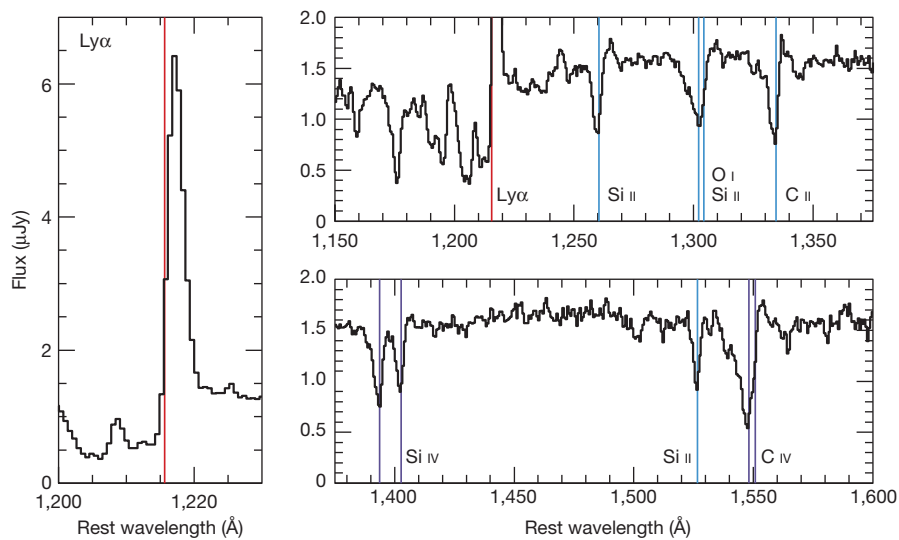

**Figure 3 | A moderately low-resolution rest-frame ultraviolet spectrum of a typical star-forming galaxy at redshift $z \approx 2$–$3$.** This spectrum, of a galaxy at $z = 2.33$ and shown in the rest frame, represents 23 h of integration with the Low Resolution Imaging Spectrometer (LRIS) on the 10-m Keck I telescope, and is therefore considerably higher in signal-to-noise ratio than most existing spectra of galaxies at comparable redshifts. The Lyα emission line is shown in the left panel, and the two right panels show continuum emission from hot, massive young stars, on which is superimposed resonance absorption lines from interstellar gas. Absorption is present from both low- and high-ionization transitions, and the strongest features are marked and labelled. Red lines indicate Lyα emission, blue lines indicate low ionization absorption lines, and purple lines indicate high ionization absorption lines.

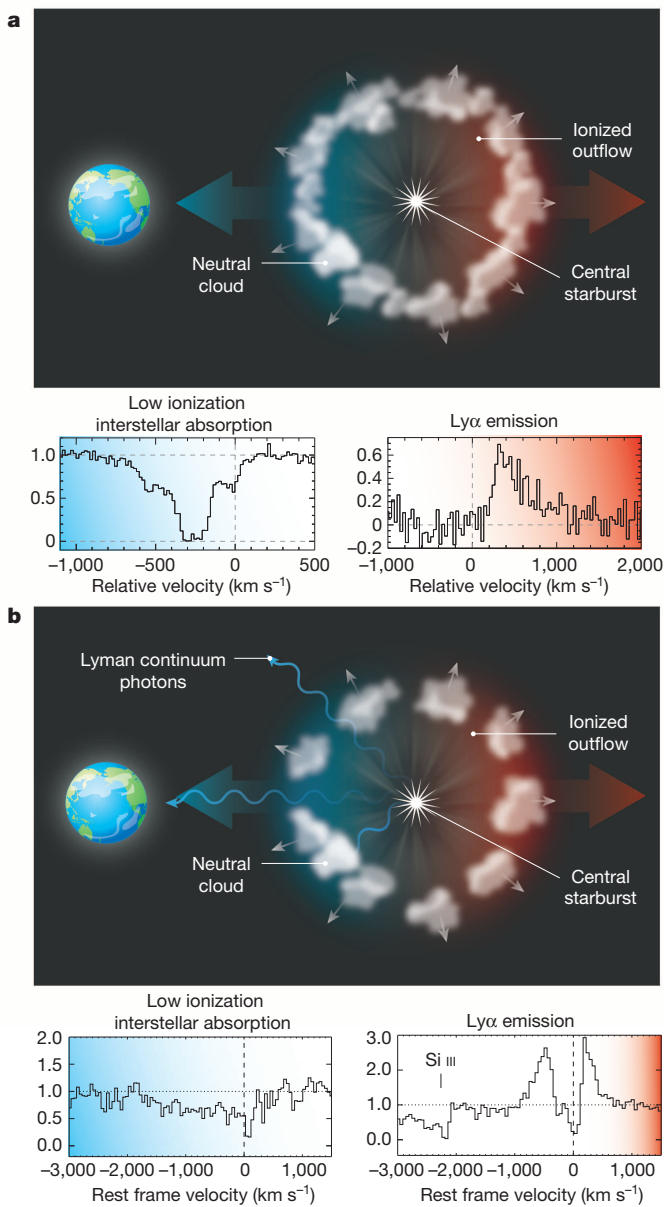

**Figure 4 | Two schematic models of a spherical galactic outflow. a**, A model in which the covering fraction of neutral hydrogen in the outflow is nearly complete. **b**, The covering fraction of neutral gas is lower, resulting in substantial residual intensity in the low ionization absorption lines, Lyα emission that is blueshifted as well as redshifted, and the potential escape of ionizing Lyman continuum photons. The spectra in **a** show low ionization absorption and Lyα emission from the $z = 2.7$ lensed galaxy MS1512-cB58[61], and the spectra in **b** show absorption and emission from the $z = 0.23$ galaxy J0921+4509[56,82]. Spectra in **a** courtesy of M. Pettini. Spectra in **b** adapted from figures 4 and 8 of ref. 56 (IOP Publishing, American Astronomical Society).

A further difficulty with observations of galactic outflows at high redshift is the low spectral resolution at which such observations are generally made, since these galaxies are typically too faint for observations at high resolution. Low resolution results in blending between absorption lines from outflowing and non-outflowing components of gas in the galaxy, with the result that the centroid of the interstellar absorption lines is a crude measurement of the outflow velocity at best; this centroid may be strongly influenced by the strength and width of absorption from gas at the redshift of the galaxy itself. This problem can be improved with the use of spectra of gravitationally lensed galaxies, in which magnification by a foreground galaxy or cluster of galaxies can result in the amplification of flux by a factor of 30 or more.

The brightness of these galaxies then allows for higher-resolution spectra in which the velocity structure of the interstellar gas can be studied in much more detail. An example of absorption lines in a gravitationally lensed galaxy is shown in Fig. 5. These lines show an absorption component at zero velocity, corresponding to the interstellar medium of the galaxy, a strong outflowing component centred at approximately $-250$ km s$^{-1}$, and a tail of outflowing gas with velocities extending to $-750$ km s$^{-1}$ (ref. 61). More recent studies of additional lensed galaxies, and of high-signal-to-noise-ratio composite spectra of large samples of galaxies, indicate that such maximum velocities are typical, with winds extending to velocities of $-800$ km s$^{-1}$ (refs 5, 62–64).

Examinations of the ultraviolet spectra of galaxies at $1.5 \lesssim z \lesssim 4$ have indicated that outflows are prevalent at these redshifts, and have provided estimates of their typical velocities, but many questions remain. One such question is whether or not the scalings of outflow properties observed in the local Universe also hold at high redshift. At $z \approx 1.4$, outflow velocity is observed to scale with star-formation rate with a comparable scaling to the local relation[4], but studies at higher redshifts have shown mixed or inconclusive results, possibly owing to the lack of dynamic range in the samples because of the inability to measure outflow velocities in very faint galaxies at high redshifts. Thus, the inclusion of faint objects in spectroscopic samples at high redshifts is key to understanding the properties of feedback in the early Universe.

### The importance of low-mass galaxies

While observations of low-mass galaxies (here defined as galaxies with stellar masses $\lesssim 10^9 M_\odot$) are necessary to an understanding of how feedback operates in the distant Universe, these objects are also important in their own right. Measurements of the rest-frame ultraviolet luminosity function of galaxies indicate that, by $z > 0.75$, the faint-end slope is steeper than in the local Universe, and it remains steep and may even increase out to the highest redshifts at which it can be measured[65–68]. Studies of samples of galaxies lensed by massive clusters also indicate that this slope remains steep down to the faintest observable magnitudes[69,70]. These results indicate that faint, low-mass galaxies host a substantial fraction of the star formation in the high-redshift Universe, while also making it clear that the determination of the contribution of faint galaxies to the global density of star formation depends on assumptions regarding the stellar populations of these faint galaxies. The metallicities, dust properties and ages of these objects are not yet well characterized[65,69].

Faint galaxies are now also being recognized as the probable key to the reionization of the Universe. Ionizing photons from the first stars and galaxies reionized the intergalactic medium, and observations now constrain the epoch at which this occurred. Spectra of quasars at $z > 6$ reveal broad, total absorption at wavelengths just short of the Lyα emission line in the spectrum of the quasar itself (the Gunn–Peterson effect), indicating the presence of neutral hydrogen in the surrounding intergalactic medium and thus suggesting the completion of reionization at $z \approx 6$ (ref. 71). Observations of the cosmic microwave background also constrain the redshift of reionization, through the increased optical depth as cosmic microwave background photons scatter off newly free electrons. Measurements of this optical depth place the redshift of reionization at $z \approx 10$–11, assuming that it was a nearly instantaneous process[72,73]; it is more likely, however, that reionization proceeded more gradually, beginning at $z > 10$ and completing at $z \approx 6$–7 (ref. 74). The production and escape of sufficient numbers of ionizing photons remains a challenge for models of reionization, with current models suggesting that large numbers of faint galaxies are required[75,76].

The optical depth of the intergalactic medium precludes detection of ionizing photons at all redshifts $z \gtrsim 4$, but studies at $z \approx 3$ now suggest that the escape fraction may be higher in faint galaxies selected by Lyα emission than in brighter, continuum-selected Lyman break galaxies (galaxies identified via broad-band imaging in filters bracketing the drop in flux at the ionization edge of hydrogen)[77–79]. Although many uncertainties remain, observations and models are beginning to build a picture of galaxies that may leak ionizing radiation. Lyman continuum photons are likely to escape from galaxies through holes in a patchy distribution of neutral hydrogen[79], and galactic winds and photoionization from massive stars are expected to be key to clearing out such channels. Galaxies locally and at $z \approx 2$ have been observed to have weak but saturated low ionization interstellar absorption lines, suggesting partial coverage of neutral gas;[56,80] they are also characterized by noticeable Lyα emission emerging blueward of the systemic velocity (see further discussion below), again characteristic of a low covering fraction of neutral hydrogen. At high redshift, these signatures are associated with intense, low-metallicity starbursts with little dust[80,81], while in the local Universe, where observations with higher spatial and spectral resolution are possible, escaping Lyman continuum emission has been directly detected from a Lyman break analogue galaxy with intense star formation and partial coverage of neutral gas, as indicated by the depths of absorption lines[82]. The general picture arising from all of these observations is of a compact, possibly low-metallicity galaxy in which strong star formation in a small volume drives powerful feedback; the combination of strong winds and intense ionizing radiation then results in incomplete coverage of neutral hydrogen, allowing some of the ionizing photons to escape. Such a scenario is illustrated in Fig. 4b.

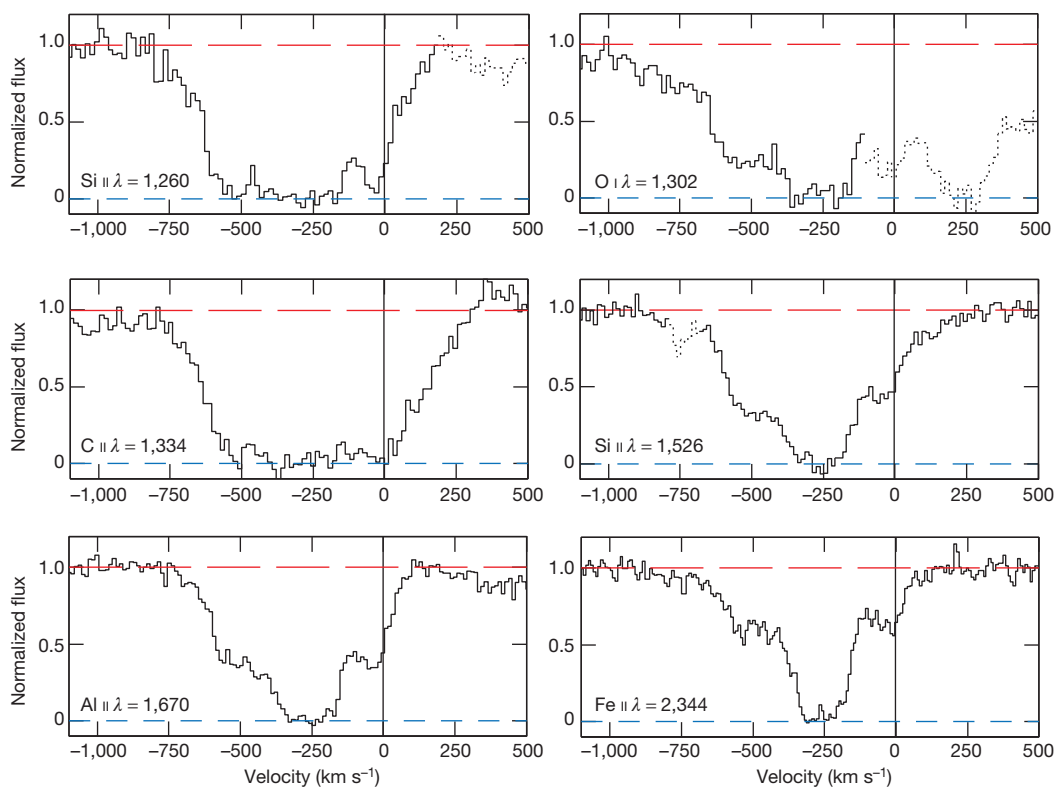

**Figure 5 | Low ionization interstellar absorption line profiles of the $z = 2.73$ gravitationally lensed galaxy MS1512-cB58.** The gas with highest optical depth has a velocity of about $250\,\mathrm{km\,s^{-1}}$, but absorption is seen over a wide range of velocities, up to $750\,\mathrm{km\,s^{-1}}$. Dotted lines show transitions other than the one labelled in each panel. Image adapted from figure 1 of ref. 61 (IOP Publishing, American Astronomical Society).

### Galactic outflows and Lyα emission

As described above, the most direct probe of outflowing gas in distant galaxies is absorption line spectroscopy. Given sufficient spectral resolution and sufficiently high signal-to-noise ratio, such spectra provide a map of the covering fraction of absorbing gas as a function of velocity, for both low and higher ionization states. Spectra with lower resolution and/or lower signal-to-noise ratio have provided valuable results through the use of long integration times or the stacking of large numbers of spectra[3,6], but the technique fundamentally requires the spectroscopic detection of continuum emission with at least a moderate signal-to-noise ratio, making its application to very faint, distant galaxies extremely challenging with current technology.

A more immediately accessible but more difficult to interpret probe of gas in galaxies at high redshifts is provided by Lyα emission. Distant galaxies can be relatively easily selected at a particular redshift by taking deep images with a narrowband filter with a central wavelength corresponding to the wavelength of Lyα at the redshift of interest[83]. Galaxies at $z \approx 2$–3 selected in this way are fainter on average than galaxies selected in typical magnitude-limited surveys at the same redshifts, with typical stellar masses of $(3$–$10) \times 10^8 M_\odot$ and little dust extinction[84–87]. Once identified, these galaxies can be studied spectroscopically; although they are faint, the combination of measurements of their Lyα profiles and systemic redshifts from rest-frame optical emission lines (shifted into the near-infrared for $z > 1.5$, which has made it difficult to obtain large samples) yields valuable constraints on the presence of galactic outflows and the covering fraction and column density of neutral hydrogen[86,88–94].

As shown in Figs 3 and 4, asymmetric, redshifted Lyα emission is a signature of outflowing gas. However, mapping the velocity profile of the Lyα emission line to the velocity structure of the outflowing gas is challenging, because the Lyα profile is affected by many other factors in addition to the gas velocity. Radiative transfer models show that the strength and velocity offset of Lyα emission depend not only on the kinematics of the outflowing gas, but on its column density, covering fraction, dust content, the angle at which the galaxy is viewed, and the presence of non-outflowing neutral hydrogen at the systemic velocity[88,95–99]. Thus, without additional constraints on these parameters, tracing the velocity structure of outflowing gas via Lyα emission is extremely difficult.

Lyα emission nevertheless provides a valuable probe of the kinematics and physical conditions of gas in galaxies that are otherwise extremely difficult to study spectroscopically. Studies of the velocity offset of Lyα emission from the systemic velocity at $z \approx 2$–3 indicate that Lyα emission is typically redshifted, even in the faintest, lowest-mass galaxies in which it has been measured (Hubble Space Telescope F814W magnitudes $m_{AB} \approx 27$, dynamical masses $M_{dyn} < 10^8 M_\odot$), indicative of the presence of outflowing gas in most faint objects studied. These studies also indicate that the velocity offset of Lyα emission increases with increasing rest-frame ultraviolet luminosity and nebular line velocity dispersion, implying that more-massive galaxies with higher star-formation rates have larger velocity offsets (see Fig. 6), and that lower-equivalent-width Lyα emission is also associated with larger velocity offsets[89,93,94].

The most obvious explanation for lower Lyα velocities in faint galaxies is that the Lyα emission is tracing a slower outflow. This is plausible; as described above, in the local Universe and to $z \approx 1.5$, the outflow velocity as traced by the centroids of absorption lines scales with both mass and star-formation rate[4,14,50,56], so a decrease in wind velocity would be unsurprising. However, this is not the only explanation for the observed trends. A further clue to the interpretation of Lyα emission in faint galaxies lies in the anti-correlation between Lyα velocity offset and equivalent width. Galaxies with stronger Lyα emission also have smaller Lyα velocity offsets, probably because an increase in the column density, covering fraction or velocity dispersion of neutral gas in the galaxy will both require that Lyα photons attain larger velocity shifts in order to escape the galaxy and increase the probability that they will be absorbed by dust during multiple scatterings or scattered beyond the spectroscopic slit[94]. Thus a decrease in the velocity offset of Lyα emission may be associated with the development of a stable, gaseous disk as galaxies grow[100].

Although Lyα emission from strongly star-forming galaxies is typically redshifted, some galaxies also show substantial Lyα emission emerging blueward of the systemic velocity[56,80,82,94]. Such blueshifted emission appears to be associated with intense, compact star formation, and appears in local galaxies with compact central sources driving high-velocity outflows[56] and at higher redshifts in galaxies with spectra indicative of low metallicity and high ionization parameters[80]. Recent observations of very faint, low-mass (approximately $(10^6-10^9) M_\odot$), lensed galaxies at $z \approx 2$ also indicate that a low metallicity and high ionization state are characteristic of low-mass galaxies, although the outflow properties of these lensed galaxies have not yet been constrained[81]. The implication (inferred also from weak but saturated low ionization absorption lines and stronger high ionization lines) is that much of the outflowing gas in such low-mass, low-metallicity galaxies is highly ionized, allowing Lyα emission to emerge with relatively little scattering by neutral hydrogen[56,80]. An absence of neutral hydrogen is also required for the escape of Lyman continuum emission, implying that galaxies with a relatively high fraction of blueshifted Lyα emission may be good candidates for Lyman continuum emission[94]; indeed, Lyman continuum photons have recently been detected from one such local galaxy[82]. Low-mass galaxies with compact star formation resulting in a highly ionized outflow may thus be the best candidates for the source of the bulk of the photons responsible for the reionization of the Universe.

## The next steps

Spectroscopic observations have now established the near ubiquity of gaseous outflows in star-forming galaxies at high redshifts, but many questions remain. There are currently very few constraints on the properties and prevalence of feedback in faint (optical $m_{AB} \gtrsim 25.5$), low-mass galaxies in the early Universe, and in part because such feedback may clear channels for the escape of the energetic photons needed to reionize the Universe, such constraints are badly needed. Direct determinations of outflow velocities in faint objects require absorption line spectroscopy; such determinations will come from long integrations, stacking very large samples, and observations of lensed galaxies. All of this work is in progress, and results will come in the next few years. Multi-object near-infrared spectrographs have already enabled systemic redshifts to be determined for larger samples of fainter galaxies than previously possible, and larger, deeper samples will constrain the physical conditions in these galaxies, relating the kinematics and covering fraction of outflowing neutral gas to the physical conditions in star-forming regions.

The next breakthroughs will be provided by future facilities. Upcoming 30-m-class telescopes (the Giant Magellan Telescope, the Thirty Meter Telescope, and the Extremely Large Telescope) will enable rest-frame ultraviolet absorption line spectroscopy of fainter galaxies at higher resolution and extending to higher redshifts; such higher-resolution observations will provide maps of the covering fraction of outflowing gas as a function of velocity. While these observations will still be limited by spectral resolution and signal-to-noise ratio, especially for fainter objects, they will undoubtedly extend our quantitative knowledge of galactic outflows at high redshift to include a much wider range of galaxy properties than is currently possible. The James Webb Space Telescope will enable near-infrared spectroscopy of galaxies at $z \gtrsim 3.5$ (the redshift at which most of the strong rest-frame optical emission lines shift beyond the ground-based atmospheric windows), allowing the determination of systemic redshifts and physical conditions in much more distant galaxies. In combination, these large ground- and space-based observatories will transform our view of the properties, prevalence, mechanisms and impact of feedback in the distant Universe.

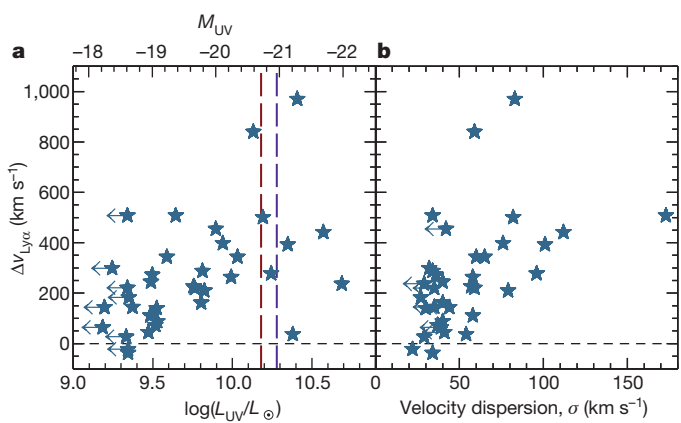

**Figure 6 | The velocity offset of Lyα emission as a function of ultraviolet luminosity and velocity dispersion, for $z \approx 2-3$ galaxies selected by strong Lyα emission.** **a**, The velocity offset of Lyα emission from the systemic velocity versus rest-frame ultraviolet luminosity, shown in units of the solar luminosity on the bottom axis and in absolute ultraviolet magnitude on the top axis. The dashed vertical lines indicate the characteristic luminosity $L^\star$ of galaxies at $z \approx 2$ (red) and $z \approx 3$ (purple)[65]. **b**, The velocity offset of Lyα emission from the systemic velocity versus velocity dispersion, for the same sample of galaxies. The velocity dispersion is measured from the width of nebular emission lines, and is an indication of the depth of the gravitational potential well. Data in this figure are taken from figures 4 and 5 of ref. 94 (IOP Publishing, American Astronomical Society).

**Acknowledgements** I thank M. Pettini for comments, suggestions, and assistance with figures, and C. Steidel, M. Pettini, A. Shapley, N. Reddy and C. Martin for many discussions and collaborations.

**Author Information** Reprints and permissions information is available at www.nature.com/reprints. The author declares no competing financial interests. Readers are welcome to comment on the online version of the paper. Correspondence and requests for materials should be addressed to the author (erbd@uwm.edu).